\documentclass[10pt] {article}
\usepackage[dvips]{graphics}
\title{Diffractive production at high energies \\ in the Miettinen--Pumplin model}
\author{S.~Sapeta\\ \\
\textit{M. Smoluchowski Institute of Physics}\\ 
\textit{Jagellonian Univeristy, Cracow
\footnote{Address: Reymonta 4, 30-059 Krakow, Poland; e-mail: sapeta@th.if.uj.edu.pl}}}
\date{}

\begin{document} 
\maketitle

\begin{abstract}
The model of soft diffractive dissociation proposed some time ago by Miettinen and Pumplin is shown to describe correctly the data at FERMILAB energies. 
The comparison with Goulianos model of Renor
malized Pomeron Flux is 
also presented.
\end{abstract}

\indent

The Regge description of diffractive dissociation \cite{Ingelman:1984ns,Kaidalov:jz}
at high energies encounters a serious problem related to unitarity. When the parameters determined below 100 GeV center of mass energy are applied to FERMILAB energies, the estimated cross section for diffraction dissociation exceeds total cross section in blatant violation unitarity and the data. To save the Regge picture K.Goulianos proposed to renormalize the pomeron flux in a way which restores unitarity and the agreement with the data \cite{Goulianos:wy}. 

The Goulianos prescription, although very elegant and effective, in not easily justifiable from theory. Therefore it seems interesting to consider other possibilities. In the present paper the high-energy data were analyzed using the parton model of Miettinen and Pumplin \cite{Miettinen:1978jb} which was shown to work very well at ISR energies. The model, based on the Good-Walker picture \cite{Good:1960ba} of diffractive processes satisfies, by construction, all unitarity constraints. We have found that a straightforward extrapolation of the Miettinen and Pumplin model to the FERMILAB energies describes very well the data. No further adjustments are necessary. One may thus conclude that the Miettinen and Pumplin description of diffractive processes provides a correct framework for analysis of the present 
experimental results.

In the Miettinen  and Pumplin model, following \cite{Good:1960ba}, the state of the incident hadron is expanded into a superposition of "diffractive" states

\begin{equation}
\mid B\rangle=\sum_{k}C_k\mid\psi_k \rangle, \label{expansion}
\end {equation}
which are eigenstates of the scattering operator
\begin{equation}
ImT\mid\psi_k \rangle=t_k\mid\psi_k \rangle.
\end {equation}
If the different eigenstates are absorbed by the target with different 
intensity, the outgoing state is no longer $\mid B \rangle$ and inelastic 
production of particles takes place. The relevant formulae for the cross sections take the form \cite{Good:1960ba}

\begin{eqnarray}
\frac{d\sigma_{el}}{d^2b }=\mid\langle B\mid ImT \mid B \rangle\mid^2
=\mid\sum_k \mid C_k \mid ^2 t_k \mid^2
=\langle t \rangle^2 \label{elb},
\end{eqnarray}
\begin{eqnarray}
\frac{d\sigma_{tot}}{d^2b}=2\langle t \rangle, \label{totb}
\end{eqnarray}
\begin{eqnarray}
\frac{d\sigma_{diff}}{d^2b}= \sum_k \mid \langle \psi_k \mid ImT \mid B \rangle - \frac{d\sigma_{el}}{d^2b} =\langle t^2 \rangle-\langle t \rangle^2 \label{difb}.
\end{eqnarray}

The basic assumption of Miettinen and Pumplin is that the eigenstates of diffraction are parton states \cite{VanHove:1976fi}

\begin{equation}
\mid \psi_k \rangle\equiv \mid \vec{b_1},...,\vec{b_N},y_1,...,y_N \rangle, \label{as}
\end {equation}
were $N$ is the number of partons; $y_i$ is the rapidity of parton $i$ and $\vec{b_i}$ 
is the impact parameter of parton $i$ relative to the impact parameter of the 
incident particle.\\
Therefore, equation  (\ref{expansion}) takes the form

\begin{equation}
\mid B\rangle=\sum_{N=0}^{\infty}\int \prod_{i=1}^{N} d^2\vec{b_i}dy_i  C_N(\vec{b_1},...,\vec{b_N},y_1,...,y_N) \mid \vec{b_1},...,\vec{b_N},
\ y_1,...,y_N \rangle.
\end{equation}
The probability $\mid C_N \mid^2$ associated with $N$ partons, which are assumed 
to be independent, is given by Poisson distribution with mean number $G^2$
\begin{equation}
\mid C_N(\vec{b_1},...,\vec{b_N}, y_1,...,y_N)\mid^2 = e^{-G^2} \frac{G^{2N}}{N!}\prod_{i=1}^{N} \mid C(\vec{b_i}, y_i)\mid^2 .
\end{equation}
To specify the eigenvalues $t_k$, Miettinen and Pumplin assumed that partons interact independently with the target. This implies that if the probability for a parton $i$ to interact is denoted $\tau_i$, the probability that none of N partons interacts is $\prod^N_{i=1}(1-\tau_i)$, hence the probability for anyone of them to interact is $1-\prod^N_{i=1}(1-\tau_i)$.\\
Miettinen and Pumplin took $\mid C(\vec{b_i}, y_i)\mid^2$ and $\tau_i(b_i,y_i)$ in a form
\begin{equation}
\mid C(b_i,y_i)\mid^2 = \frac{1}{2\pi\beta\lambda}
exp\left(-\frac{\mid y_i\mid}{\lambda}-\frac{b_i^2}{\beta}\right),
\end{equation}

\begin{equation}
\tau_i(b_i,y_i)= Aexp\left(-\frac{\mid y_i\mid}{\alpha}-\frac{b_i^2}{\gamma}\right),
\end{equation}
with

\begin{equation}
A=1 \qquad  \frac{\alpha}{\lambda}=2.0 \qquad	 \frac{\gamma}{\beta}=2.0
\end{equation}
Taking all this into consideration, the total, elastic and diffractive cross sections are given by \cite{Miettinen:1978jb}

\begin{eqnarray}
\frac{d\sigma_{tot}}{d^2b}=2\left(1-exp\left(-G^2\frac{4}{9}e^{-\frac{1}{3}\cdot\frac{b^2}{\beta}}\right)\right),\label{tot}
\end{eqnarray}

\begin{eqnarray}
\frac{d\sigma_{el}}{d^2b}=\left(1-exp\left(-G^2\frac{4}{9}e^{-\frac{1}{3}\cdot\frac{b^2}{\beta}}\right)\right)^2,\label{elastic}
\end{eqnarray}

\begin{eqnarray}
\frac{d\sigma_{diff}}{d^2b}=exp\left(-2G^2 \frac{4}{9}e^{-\frac{1}{3}\cdot\frac{b^2}{\beta}}\right) \left(exp\left(G^2\frac{1}{4}e^{-\frac{1}{2}\cdot\frac{b^2}{\beta}}\right)-1\right). \label{sigma_b}
\end{eqnarray}
\begin{figure}[tb]
	\begin{center}
		\rotatebox{270}{\scalebox{0.4}{\includegraphics{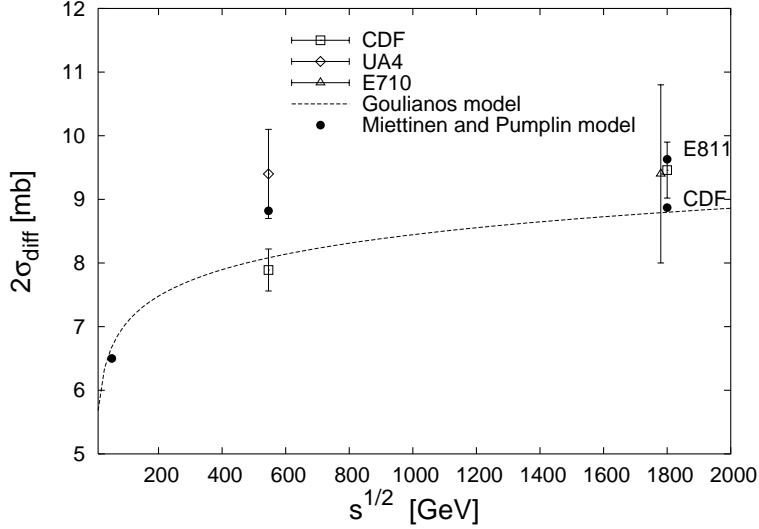}}}
	\end{center}
\caption{Total $pp$ single diffraction cross section data compared with predictions based on Miettinen and Pumplin model and Goulianos model.}
	\label{th-exp}
\end{figure}
As we see, the model depends on two parameters $G^2$ and $\beta[fm^2]$. Those 
parameters can be determined for a given energy $\sqrt{s}$ from experimental 
data for $\sigma_{tot}$ and $\sigma_{el}$ using (\ref{tot}) and (\ref{elastic}).

In 1978 Miettinen and Pumplin performed calculations for two colliding 
protons with the center of mass energy $\sqrt{s}=53GeV$. They obtained the 
result for $\sigma_{SD}$ which were in good agreement with the data.

We have applied the Miettinen and Pumplin model to energies $\sqrt{s}=546GeV$ and 
$\sqrt{s}=1800GeV$ using the CDF \cite{Abe:1993xy} and E811 \cite{Avila:1998ej} data for $\sigma_{tot}$ and $\sigma_{el}$.\\
The obtained values of $G^2$ and $\beta[fm^2]$, presented in table \ref{tab1}, allow to calculate the cross section for 
single diffractive production. 
\begin{table}[b]
\begin{center}
\begin{tabular}{|c|c|c|c|c|c|c|} \hline
\small{references} & \small{$\sqrt{s}${[GeV]}} & \small{$\sigma_{tot}[mb]$} & \small{$\sigma_{el}[mb]$}&\small{$G^2$}&\small{$\beta[fm^{2}]$}&\small{$2\sigma_{diff}[mb]$}\\ \hline
\cite{Miettinen:1978jb}   & 53   & 43             &   8.7           & 2.91   & 0.235 & 6.51 \\ \hline
CDF \cite{Abe:1993xy} & 546  & 61.26$\pm$0.93 & 12.87$\pm$0.30  & 3.12   & 0.319 & 8.82 \\ \hline
E811 \cite{Avila:1998ej}
& 1800 & 71.71$\pm$2.02     &  15.79$\pm$0.87 & 3.38   & 0.351 & 9.63 \\ \hline
CDF \cite{Abe:1993xy}& 1800 & 80.03$\pm$2.24 & 19.70$\pm$0.85  & 4.20   & 0.337 & 8.87 \\ \hline
\end{tabular}
\end{center}
\caption{Total and elastic cross sections together with the obtained values of $G^2$ and $\beta[fm^2]$. The last column includes calculated values of diffractive cross sections.}
\label{tab1}
\end{table}
Figure \ref{th-exp} shows the results compared with 
experiments and with the Goulianos model. To take into account the beam and the target 
dissociation, $\sigma_{SD}$ is multiplied by the factor of two. 

The two values for $\sqrt{s}=1800GeV$ are consequence of two different results 
for $\sigma_{tot}$ and $\sigma_{el}$ measured by CDF  and E811. We 
see from the table \ref{tab1} that Miettinen and Pumplin model is valid in a considerable range of energies. 
Similarly to the Goulianos model it predicts a slow rise of $\sigma_{SD}$ with 
$\sqrt{s}$. It gives, however the values which are a little higher.

\begin{figure}[htb]
	\begin{center}
		\rotatebox{270}{\scalebox{0.4}{\includegraphics{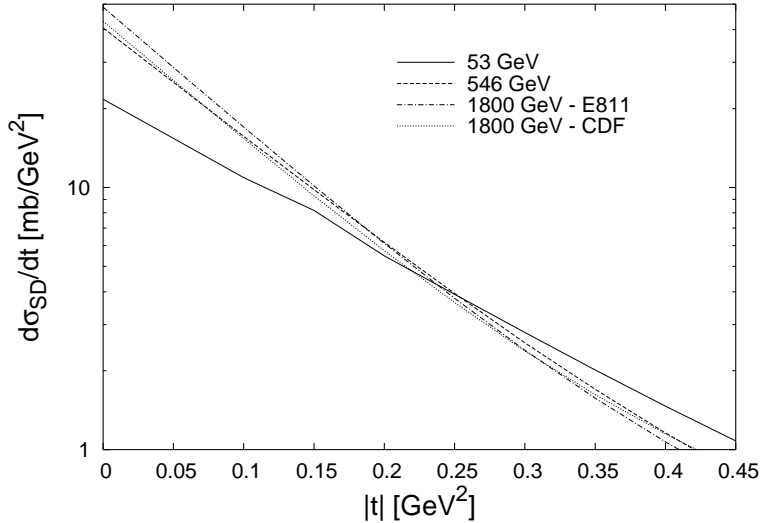}}}
	\end{center}
\caption{The momentum-transfer dependence of a beam dissociation obtained within Miettinen and Pumplin model.}
	\label{s_t}
\end{figure}
The dependence of diffractive dissociation cross section, shown in figure \ref{s_t}, on
the momentum transfer $t$  is obtained by applying Fourier transform to (\ref{sigma_b}). One sees that the slope increases with energy. The calculated 
values of the slope for energy $\sqrt{s}=1800GeV$ are $9.9$ (CDF) and $10.2$ (E811), 
which is consistent with measurement of E710 \cite{Amos:1992jw} i.e. $10.5 \pm 1.8$. We have also checked that the elastic slopes, calculated from (\ref{elastic}), are consistent with existing experimental data.

In conclusion, we have shown that the Miettinen and 
Pumplin model correctly describes diffraction dissociation  in hadron-hadron collisions with the energies of the order of TeV. 
Calculated values of $\sigma_{SD}$ are  in reasonable agreement with experimental 
data. Moreover, the dependence on energy is similar to that calculated by 
Goulianos within his model of renormalized Pomeron flux. 
\paragraph{Acknowledgements\\ \\}

I would like to thank A.Bialas for suggesting this investigation and helpful remarks.

\end{document}